\documentclass[preprint,5p,number,twocolumn]{elsarticle}
\usepackage{graphicx,epsfig}
\usepackage{dcolumn}
\usepackage{bm}
\usepackage{psfrag}
\usepackage{plain}
\usepackage{subfig}
\usepackage{tabularx}
\usepackage[latin1]{inputenc}
\usepackage{amsmath}
\usepackage{amssymb}
\usepackage{graphics}
\usepackage{graphicx}
\usepackage{epsfig}

\usepackage{amssymb}
\usepackage{amsthm}

\journal{Physica A}

\begin{document}

\begin{frontmatter}


\title{Linear stability analysis of transverse dunes}


\author[ufc]{Hygor P. M. Melo\corref{hp}}\ead{hygor@fisica.ufc.br}\author[ufc]{Eric J. R. Parteli}\ead{parteli@fisica.ufc.br}\author[ufc]{Jos\'e S. Andrade Jr.}\ead{soares@fisica.ufc.br}\author[ufc,eth]{Hans J. Herrmann}\ead{hans@ifb.baug.ethz.ch}
\address[ufc]{Departamento de F\'{\i}sica, Universidade Federal do Cear\'a - 60455-760, Fortaleza, CE, Brasil.}
\address[eth]{Institut f\"ur Baustoffe IfB, ETH H\"onggerberg, HIF E 12, CH-8093, Z\"urich, Switzerland.}


\begin{abstract}
Sand-moving winds blowing from a constant direction in an area of high sand availability form transverse dunes, which have a fixed profile in the direction orthogonal to the wind. Here we show, by means of a linear stability analysis, that transverse dunes are intrinsically unstable. Any along-axis perturbation on a transverse dune amplify in the course of dune migration due to the combined effect of two main factors, namely: the lateral transport through avalanches along the dune's slip-face, and the scaling of dune migration velocity with the inverse of the dune height. Our calculations provide a quantitative explanation for recent observations from experiments and numerical simulations, which showed that transverse dunes moving on the bedrock cannot exist in a stable form and decay into a chain of crescent-shaped barchans.

\end{abstract}

\begin{keyword}
Sand dunes \sep Pattern formation \sep Wind erosion \sep Granular matter \sep Transverse instability
\PACS{45.70.Qj, 45.70.-n, 92.40.Gc, 92.60.Gn}


\end{keyword}

\end{frontmatter}




\section{Introduction}

Sand dunes are widespread on Earth deserts and coasts, and are also found on Mars, Venus, Titan and even on the bottom of rivers \citep{Bagnold_1941,Cutts_and_Smith_1973,McCauley_1973,Greeley_et_al_1992,Lorenz_et_al_2006,Fourriere_et_al_2010,Bourke_et_al_2010}. Since the pioneering work by \cite{Bagnold_1941}, insights from field and experimental works \citep{Pye_and_Tsoar_1991,Wiggs_2001,Hersen_et_al_2002,Elbelrhiti_et_al_2005,Livingstone_et_al_2007,Andreotti_et_al_2009,Andreotti_et_al_2010,Reffet_et_al_2010}, as well as numerical simulations \citep{Werner_1995,Nishimori_et_al_1998,Sauermann_et_al_2001,Kroy_et_al_2002,Herrmann_et_al_2008,Narteau_et_al_2009,Parteli_et_al_2009}, have steadily refined our knowledge on the physics of sand transport and dune formation.

\begin{figure*}[htpb]
\begin{center}
\includegraphics[width=0.67 \textwidth]{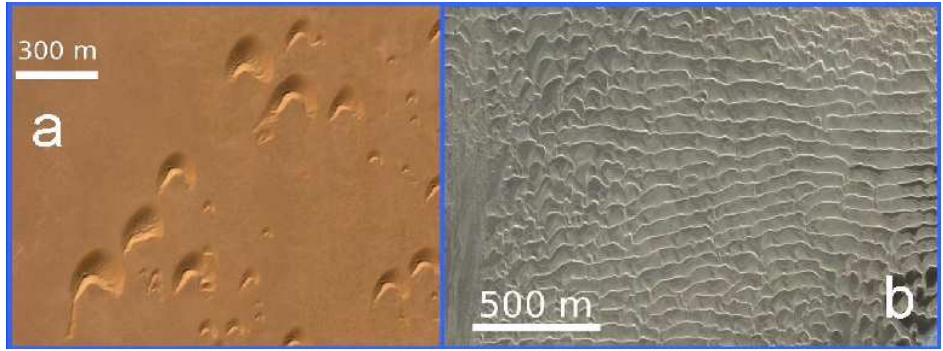}
\caption{{\bf{a.}} Barchan dunes in Westsahara, near $26.51^{\circ}$N, $13.21^{\circ}$W. {\bf{b.}} Transverse dunes in Baja California, near $28.06^{\circ}$N, $114.05^{\circ}$W. In both images, the orientation of the dunes show that the wind is roughly undirectional and blows from the top. Images courtesy of Labomar. } 
\label{fig:dune_shapes}
\end{center}
\end{figure*}

Dunes form wherever sand is exposed to a wind that is strong enough to put grains into saltation --- which consists of grains jumping in nearly ballistic trajectories and ejecting new particles upon collision with the sand bed \citep{Bagnold_1941,Sauermann_et_al_2001,Andreotti_2004,Almeida_et_al_2006,Almeida_et_al_2008,Kok_and_Renno_2009}. The main factors controlling dune morphology are the wind directionality and the amount of loose sand available for transport \citep{Wasson_and_Hyde_1983}. While elongating, longitudinal seif dunes and accumulating star dunes develop under bi- and multidirectional wind regimes, respectively \citep{Pye_and_Tsoar_1991}, in areas where the wind direction is nearly constant, two types of migrating dune may occur: 
\begin{itemize}
\item {\em{barchan dunes}}, which have a crescent shape and two limbs pointing in the migration direction (Fig.~\ref{fig:dune_shapes}a); they occur in areas where the amount of sand is not sufficient to cover the ground \citep{Finkel_1959,Long_and_Sharp_1964,Embabi_and_Ashour_1993,Hesp_and_Hastings_1998,Sauermann_et_al_2000,Sauermann_et_al_2003,Bourke_and_Goudie_2009};
\item {\em{transverse dunes}}, which propagate with nearly invariant profile orthogonally to a fixed wind direction (Fig.~\ref{fig:dune_shapes}b); they form when the amount of sand on the ground is high \citep{Mulligan_1988,McKenna_Neuman_et_al_2000,Walker_and_Nickling_2002,Parteli_et_al_2006}. The transverse dune is the simplest and best-understood type of dune, and occurs on all planetary bodies where dunes have been detected \citep{Bourke_et_al_2010}. 
\end{itemize}

\begin{figure}[htpb]
\begin{center}
\includegraphics[width=0.9 \columnwidth]{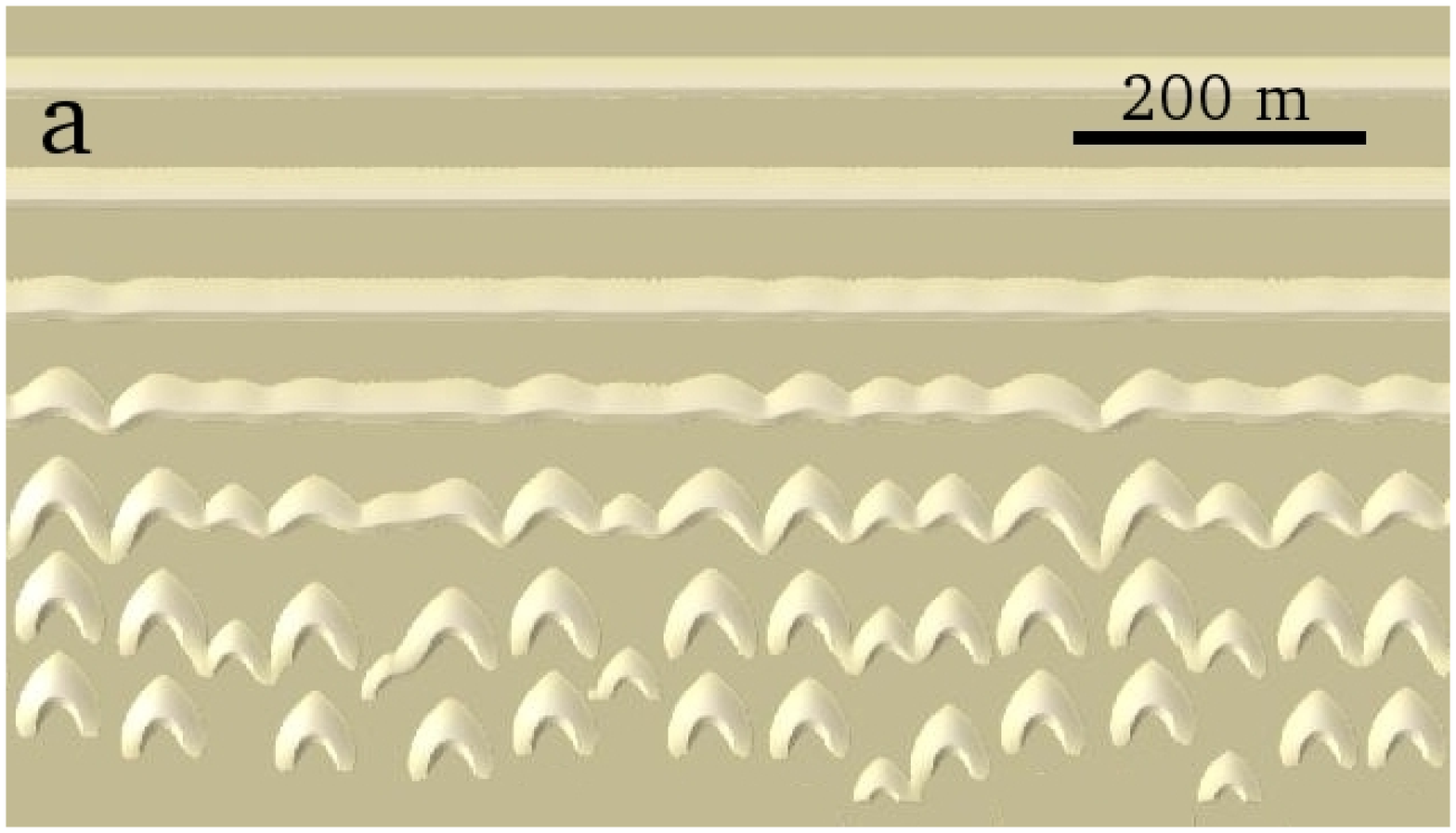}
\includegraphics[width=0.9 \columnwidth]{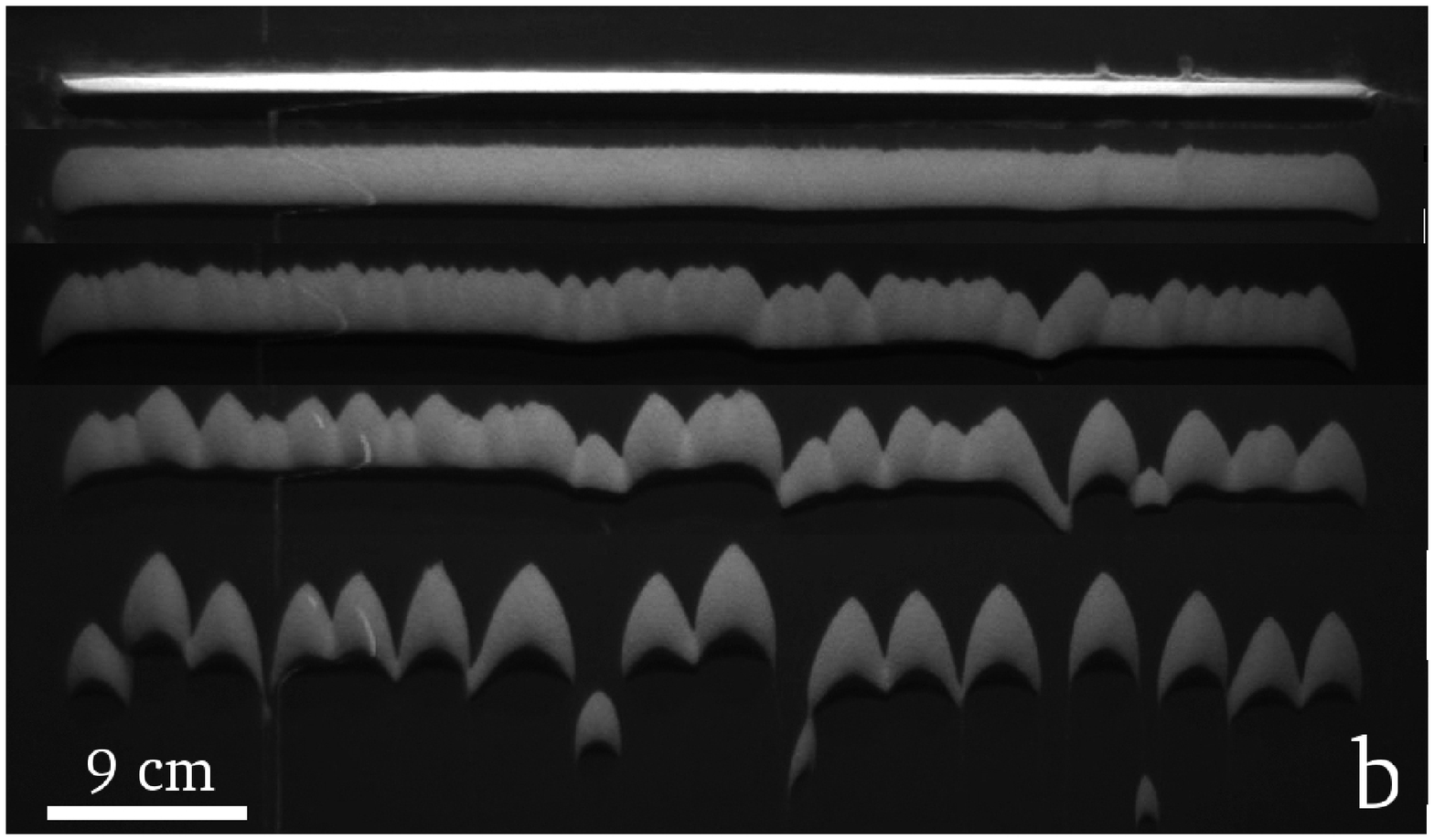}
\caption{Evidence for the transverse instability of dunes from: {\bf{(a)}} numerical simulations \citep{Parteli_et_al_2011} --- spatio-temporal sketch of the profile of a transverse dune at different times, which is evolving under constant wind direction. Wind blows from the top. The dune is unstable and decays after some distance of migration into a chain of barchans; {\bf{(b)}} water-tank experiments \citep{Reffet_et_al_2010}: snapshots of one experimental realization of a transverse dune evolving under constant water stream (flow direction is from the top). An initial sand barrier evolves into a transverse dune of width $\sim 3$~cm, which then decays into a chain of barchans. Images credit: Sylvain Courrech du Pont - Laboratoire MSC (Mati\`ere et Syst\`emes Complexes) - UMR 7057 - Universit\'e Paris Diderot, Paris, France.}  
\label{fig:transverse_instability}
\end{center}
\end{figure}

As a matter of fact, barchans are observed in bedrock areas of Earth's desert and coastal environments \citep{Lancaster_1995} or on the floor of Martian craters \citep{Bourke_et_al_2004} where sand availability is low. Water tank experiments of dune formation under unidirectional stream also produce barchan dunes if the sand cover is incipient or there is no input of sand \citep{Katsuki_et_al_2005,Reffet_et_al_2010}. These experiments, as well as numerical simulations \citep{Duran_et_al_2010,Luna_et_al_2011,Parteli_et_al_2011}, suggest that transverse dunes cannot exist in a stable form on the bedrock: a transverse sand ridge evolving on the bare ground under unidirectional wind or water stream decays into a chain of barchans after some distance of migration (cf. Fig.~\ref{fig:transverse_instability}a,b). Evidence for this instability of transverse dunes has been also reported from field observations \citep{Kocurek_et_al_1992,Luna_et_al_2011}: transverse dunes emerging in an area of dense sand cover (e.g. a flat sand beach in a coastal area) decay into barchans after migrating a certain distance on the bedrock downwind. Therefore, this type of dune instability clearly plays a crucial role for the genesis and dynamics of barchan dune fields.

Previous studies of dune genesis have focused on the growth of transverse dunes from a sand hill or a flat sand bed subjected to a wind of constant direction \citep{Kennedy_1969,Smith_1970,Richards_1980,McLean_1990,Stam_1997,van_Dijk_et_al_1999,Kroy_et_al_2002,Andreotti_et_al_2002,Schwaemmle_and_Herrmann_2004,Fourriere_et_al_2010}. However, the stability of the classical transverse dune shape has remained an open issue for several decades \citep{Reffet_et_al_2010}. A quantitative study of the long-term evolution of a transverse dune was performed recently by means of numerical simulations using a model for sand transport in three dimensions \citep{Parteli_et_al_2011}. More precisely, it was shown that transverse dunes are unstable with respect to any along-axis perturbation in their profile, regardless of wind strength, dune size or whether the perturbation is random or periodic \citep{Parteli_et_al_2011}. No matter the type of initial perturbation, the transverse dune breaks into a chain of barchans, each displaying an average cross-wind width of the order of 10 times the original dune height. Indeed, the total time of the decay process depends only on the dune height and the magnitude of the initial perturbations, being independent of the wavelength of the perturbation \citep{Parteli_et_al_2011}. 

The question which we want to address in the present paper is the following: why does an along-axis perturbation in the dune profile grow? Here we want to address this issue in a quantitative sense performing a linear stability analysis of the transverse dune. While numerical simulations can be used to make quantitative predictions of the long-term behaviour of a transverse dune \citep{Parteli_et_al_2011}, the linear stability analysis refers to the early stage of the perturbation growth, when the magnitude of the perturbation is small \citep{Anderson_1987,Andreotti_et_al_2002}. We consider an infinitely long transverse dune to which a harmonic modulation of small magnitude is added. The time evolution of this perturbation is calculated by considering a strictly unidirectional wind regime.

This paper is organized as follows. In the Section 2 we give a brief review on the physics of sand transport and the formation of transverse dunes. The linear stability analysis of the transverse dune is the subject of Section 3. Finally, conclusions are presented in Section 4.

\section{\label{sec:dune_physics}The physics of dune formation}

\subsection{Formation of transverse dunes from a longitudinal sand-wave instability}

A sand plane subjected to a wind of sufficient strength is unstable and gives rise to a chain of transverse dunes \citep{Kennedy_1969,Smith_1970,Richards_1980,McLean_1990,Fowler_2001,Fowler_2002,Andreotti_et_al_2002,Schwaemmle_and_Herrmann_2004}. This instability is of hydrodynamic origin. A bump or small hill poses an obstacle to the wind, thus producing an upward force that deflects the air flow approaching the bump from the upwind. At the upper portion of the bump, a downward force keeps the flow attached to the surface, thus pushing the flow streamlines closer to each other and enhancing the air shear stress on the top of the hill. However, the maximum of the shear stress is not exactly at the crest, rather it is shifted upwind with respect to the bump's profile \citep{Stam_1997,Kroy_et_al_2002,Fourriere_et_al_2010}. So, if the flux reacted without delay to a variation in wind velocity, then maximal erosion would always take place upwind of the crest; sand would be then always deposited on the crest, and the bump would grow.

\begin{figure*}[htpb]
\begin{center}
\includegraphics[width=0.6 \textwidth]{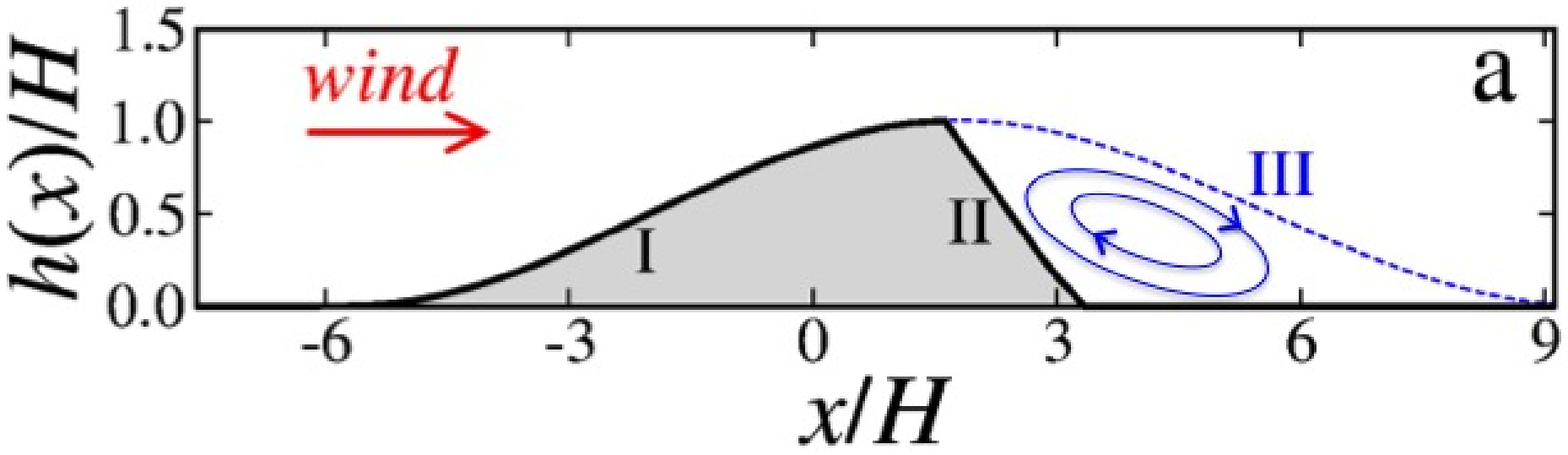}
\includegraphics[width=0.2 \textwidth]{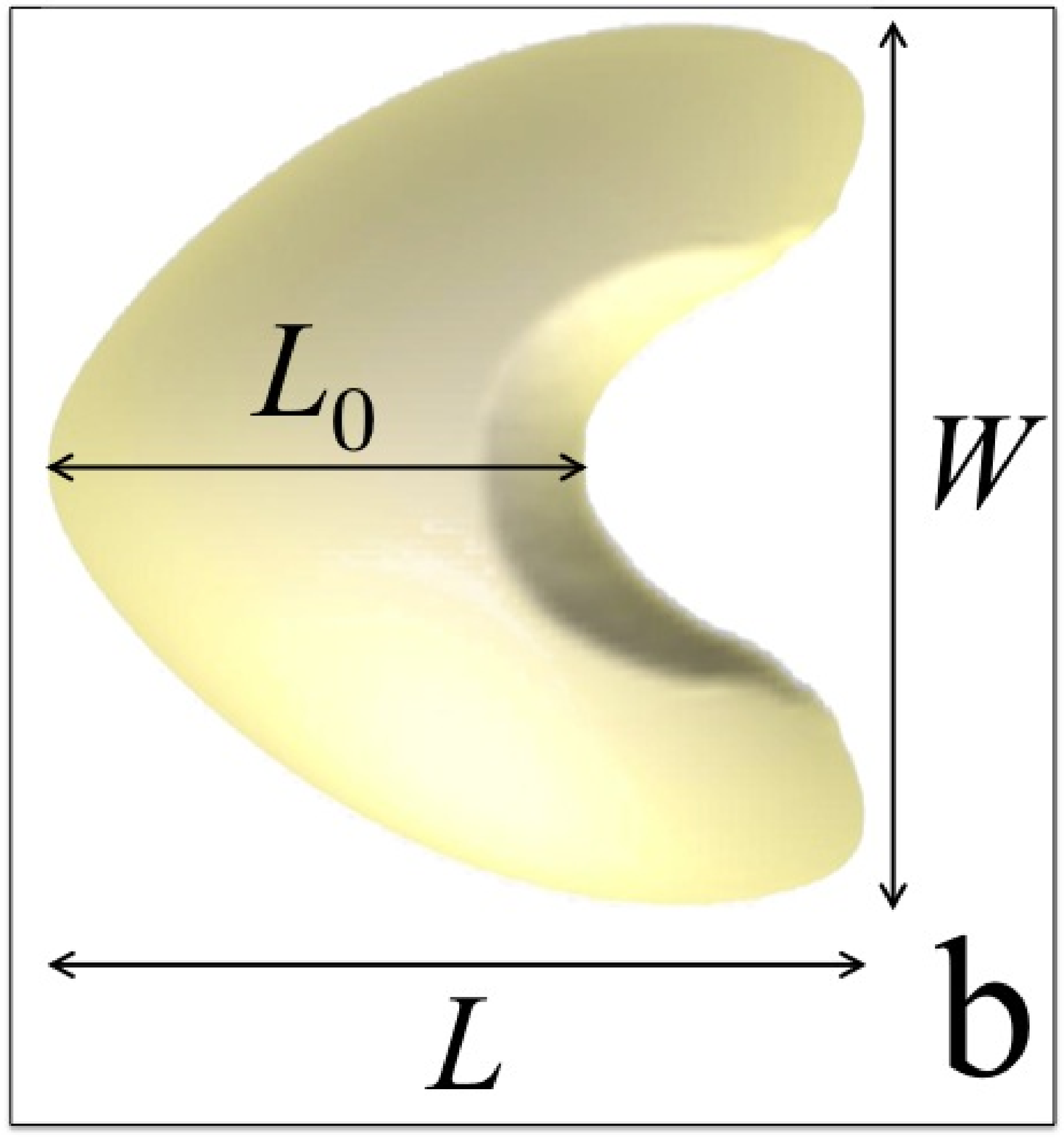}
\caption{{\bf{(a)}} Schematic diagram illustrating the longitudinal profile of a transverse dune (full line) and the corresponding separation streamline of the wind flow at the lee (dashed line). Inside the so-called ``separation bubble'', a zone of recirculating flow develops, within which net transport vanishes. Both the downwind position $x$ and the dune profile $h(x)$ are rescaled by the height of the dune ($H$). {\bf{(b)}} Sketch of a barchan dune of length $L$ and cross-wind width $W$. The central slice of the barchan mimicks the longitudinal profile of a transverse dune of width $L_0$.}  
\label{fig:dune_sketch}
\end{center}
\end{figure*}

Indeed, there is a transient length needed for the saltation flux to adapt to a change in wind speed. This length-scale is the so-called saturation length \citep{Sauermann_et_al_2001,Andreotti_et_al_2010},
\begin{equation}
L_{\mathrm{sat}} \approx 2\,{\frac{{\rho}_{\mathrm{grains}}}{{\rho}_{\mathrm{air}}}}d,
\end{equation}
where ${\rho}_{\mathrm{grains}}$ and ${\rho}_{\mathrm{air}}$ are the density of the grains and of the air, respectively, and $d$ is the grain diameter. For quartz particles (${\rho}_{\mathrm{grains}} = 2650$ kg$/$m$^3$) of average diameter $d = 250$ ${\mu}$m \citep{Bagnold_1941}, $L_{\mathrm{sat}}$ for saltation under Earth conditions (${\rho}_{\mathrm{air}} = 1.225$ kg$/$m$^3$) is approximately 1~m. The bump grows only if its width is large enough such that the sand flux attains its maximal value upwind of the bump's crest. If the hill is smaller than about $20L_{\mathrm{sat}}$ \citep{Fourriere_et_al_2010}, then it is completely eroded and disappears. Conversely, a hill that is larger than this minimal size can evolve into a transverse dune. The shape of the hill becomes increasingly asymmetric as the lee slope increases due to deposition, while the windward side becomes correspondingly less steep as a result of erosion. Eventually, the slope at the downwind side of the dune becomes so large that the flow streamlines there cannot be kept attached to the surface and flow separation occurs. At the lee, a zone of recirculating flow develops which extends downwind up to a distance of about $4-8$ times the dune height \citep{Herrmann_et_al_2005,Parteli_et_al_2006}. This so-called ``separation bubble'' (c.f. Fig.~3) functions as a sand trap, since net transport there essentially vanishes \citep{Wiggs_2001}. Further, when the lee slope exceeds the angle of repose of the sand, ${\theta}_{\mathrm{c}} \approx 34^{\circ}$, the surface downwind relaxes through avalanches in the direction of the steepest descent, forming the so-called slip-face, as depicted in Fig.~\ref{fig:dune_sketch}a. A sharp brink separates the slip-face from the windward side, which has slopes of about $5^{\circ} - 10^{\circ}$ \citep{Sauermann_et_al_2000,Hesp_and_Hastings_1998}. Dune migration consists of grains climbing up the windward side through saltation, being deposited downwind of the crest and thereafter sliding down the slip-face through avalanches. 

The transverse dune migrates downwind with a velocity $v$ that is roughly inversely proportional to the dune height ($H$). In fact, the larger the size of the dune, the more time is needed for a dune to migrate a distance equivalent to its width ($L_0$). The dune turnover time may be understood as the time needed for a sand grain, once buried on the foot of the slip-face, to reappear on the foot of the dune's windward side \citep{Allen_1974,Hersen_et_al_2004,Duran_et_al_2010}.

\subsection{Barchan dunes}

In the discussion above, we have considered a sand hill that is invariant in the direction perpendicular to the wind, i.e. an infinitely long transverse sand ridge. If the initial surface is not a transverse ridge but rather a small heap approximately as long as it is wide, then the longitudinal instability leads to a barchan dune (Fig.~\ref{fig:dune_sketch}b). Due to the scaling $v \propto 1/H$ \citep{Bagnold_1941}, the central slice of the heap, which is the one with the largest height, migrates with the smallest velocity. In the opposite, the sidemost slices advance downwind the fastest, and in this manner a crescent shape is formed with two limbs pointing in the wind direction. Indeed, if the different slices moved uncoupled to each other, then they would migrate apart due to their different speeds. However, the slices are coupled due to lateral sand transport in the direction perpendicular to dune migration. This lateral transport occurs on the slip face due to gravitational downslope forces arising wherever on the slip face the slope exceeds the angle of repose, and on the windward side due to the lateral component of the wind shear stress \citep{Duran_et_al_2010}. The two factors are negligible near the crest where the lateral slope is small and so the central slice of a barchan essentially mimicks the cross section of a transverse dune of same height.

\section{Linear stability analysis of the transverse dune}

In this Section, we aim to verify the transverse instability of dunes through a linear stability analysis. In fact, since we do not know the exact analytical form of the transverse dune's cross-section in the steady state, i.e. the unperturbed solution, it is not possible to perform the full stability analysis of the three-dimensional dune. However, the problem can be simplified by analyzing the perturbation at the slip face of the dune, which is where the decisive mechanism of lateral transport occurs. 

Following the discussion of the previous section we assume that the migration velocity of a thin longitudinal slice of a transverse dune is essentially inversely proportional to its height \citep{Bagnold_1941}, or more precisely to the square root of its area. So if a perturbation in the area of the slices is applied along the transverse direction, then the smaller slices move faster, while the larger ones stay behind. In order to assure the existence of an instability, it just suffices to demonstrate that there is a lateral flux of sand from the advanced slices of the dune towards those that are behind. If this is the case, then the smaller slices will become even smaller, and the larger ones even larger, in such a manner that their relative speed difference increases, therefore enhancing the perturbation. 
 
We consider an infinite and straight transverse dune, which is moving under a constant wind that is strictly unidirectional in the $x$-direction. Let $M$ be the mass of this small longitudinal slice of this dune of width $dy$. So $M$ can be approximated as $M \approx \rho Ady$, where $\rho$ is the bulk density of the dune, taken as a constant in our calculations. Since the slip-face can be regarded as a sand trap (c.f. Section \ref{sec:dune_physics}), the flux leaving a given slice in wind direction at the lee side should be negligible. Thus, any change in the mass profile occurs due to lateral transport along the $y$ axis, i.e. due to the mass exchange between neighbouring longitudinal slices. At time $t=0$, we add to the area of the dune an infinitesimal harmonic perturbation in the direction orthogonal to the wind (the $y$-direction), such that the perturbed area profile can be written as,
\begin{equation}
A(y,t)=A_{0}[1+\epsilon\exp({\Lambda t})\cos({\omega y})],
\label{eq:Area_pert}
\end{equation}
where $A_0$ is the area of the unperturbed transverse dune, ${\omega}$ and ${\epsilon}$ are the frequency and the initial amplitude of the perturbation, respectively, and ${\Lambda}$ is the perturbation's growth rate. From Eq.~(\ref{eq:Area_pert}), it can be seen that the sign of $\Lambda$ dictates the long-time behaviour of the perturbation. If ${\Lambda} < 0$, then the perturbation decreases and the transverse dune is stable, while a positive value of ${\Lambda}$ makes the perturbation increase, which means that the transverse dune is unstable. 

\begin{figure}[htpb]
\begin{center}
\includegraphics[width=0.8 \columnwidth]{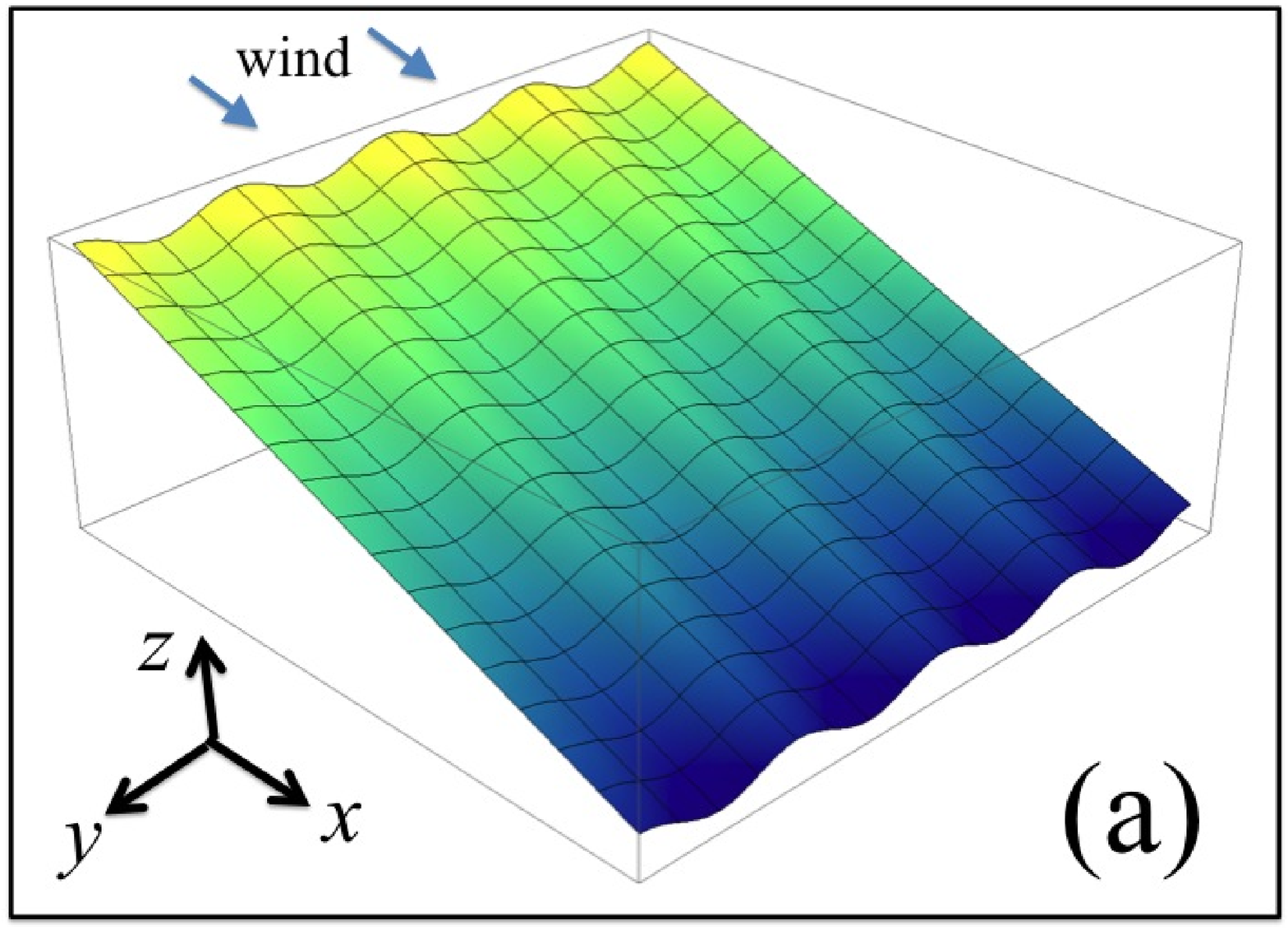}
\includegraphics[width=0.8 \columnwidth]{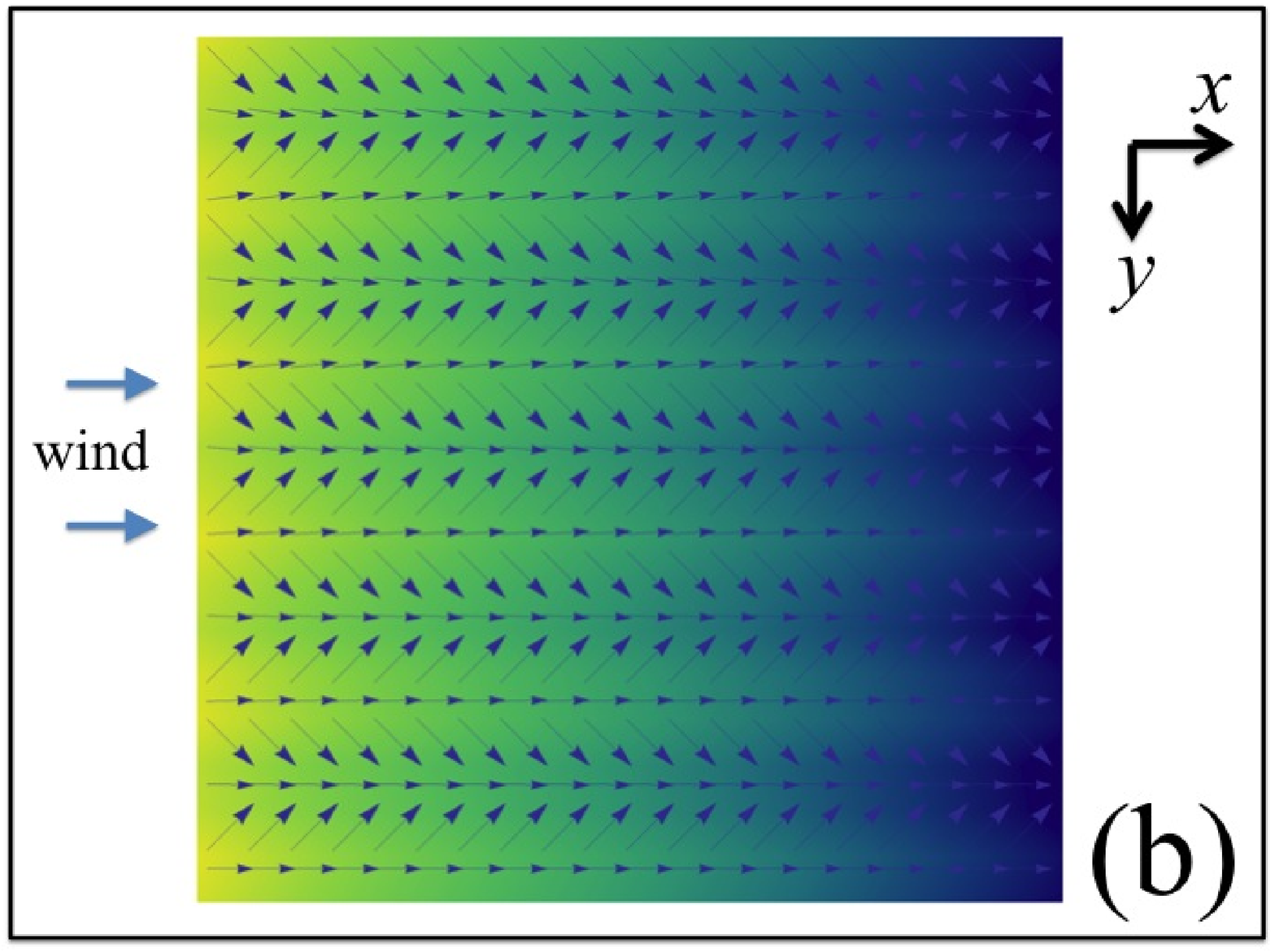}
\caption{(a) Schematic diagram showing the slip-face's surface, $\sigma(x,y)$ (defined as in Eq.~(\ref{eq:sigma})) of the transverse dune perturbed according to Eq.~(\ref{eq:Area_pert}). (b) Vector field $\hat{f}$ (Eq.~(\ref{eq:f})) of the sand flux along the perturbed transverse dune's slip-face (viewed from the top).}  
\label{fig:hygor}
\end{center}
\end{figure}

The migration velocity of a thin longitudinal slice follows the relation $v(y,t) = 2c/\sqrt{A(y,t)}$, where $c$ is a constant which depends on the shear stress. Thus, the periodic perturbation in the area profile, Eq.~(\ref{eq:Area_pert}), induces a perturbation in the velocity of the form,
\begin{equation}
v(y,t)=\frac{2c}{\sqrt{A_{0}[1+\epsilon\exp({\Lambda t})\cos{(\omega y)}]}}.
\label{eq:velx1}
\end{equation}
Since $\epsilon$ is small, we can expand the expression above in a Taylor series up to the first order in ${\epsilon}$, such that $v(y,t)$ can be approximated to,
\begin{equation}
v(y,t)=\frac{2c}{\sqrt{A_{0}}}{\left[{1-\frac{1}{2}\epsilon\exp({\Lambda t})\cos{(\omega y)}}\right]}.
\label{eq:velx2}
\end{equation}
Integration over time of Eq.(\ref{eq:velx2}) gives the position $x(y,t)$ of each longitudinal slice,
\begin{equation}
x(y,t)=x_{0}+\frac{2c}{\sqrt{A_{0}}}t-\frac{c}{\Lambda\sqrt{A_{0}}}\epsilon\exp{(\Lambda t)}\cos{(\omega y)},
\label{eq:x_y_t}
\end{equation}
where we assumed, for convenience, that the initial position of the slices is 
\begin{equation}
x(y,0)=x_{0}-\frac{c}{\Lambda\sqrt{A_{0}}}\epsilon\cos{(\omega y)}.
\label{eq:x0_pert}
\end{equation} 
In fact, choosing such an initial profile for $x(y,0)$ means adding to the initial transverse dune also a perturbation on the $xy-$plane (besides the perturbation in the height given by Eq. (\ref{eq:Area_pert})). However, Eqs.~(\ref{eq:Area_pert}) and (\ref{eq:x0_pert}) mean that the smaller slices of the initial transverse dune are slightly advanced downwind with respect to the larger ones --- this must occur also if an invariant profile ($x(y,0) = x_0$) is taken instead: due to the scaling $v \sim 1/H$, the smallest slices of a transverse dune that is perturbed according to Eq.~(\ref{eq:Area_pert}) will move in the front after a certain (infinitesimally small) amount of time, such that a perturbation in the height also gives rise to a perturbation on the $xy$-plane (Eq.~(\ref{eq:x0_pert})). In other words, rather than the specific initial profile of $x(y,0)$, it is the dynamics of lateral transport in the course of dune motion which is relevant for the stability analysis of the transverse dune.
From the profile $x(y,t)$, as shown in Fig.~\ref{fig:hygor}a, we can obtain the surface equation of the slip-face,
\begin{equation}
\begin{split}
\sigma(x,y) &= (x,y,x_{0}\tan(\theta_{c}) +v_{0}t\tan{(\theta_{c})}-x(y,t)\tan{(\theta_{c})} \\
& - \frac{c}{\Lambda\sqrt{A_{0}}}\epsilon\exp({\Lambda t})\cos{(\omega y)}\tan{(\theta_{c})}),
\label{eq:sigma}
\end{split}
\end{equation}
where $v_0 = 2c/{\sqrt{A_0}}$. Equation~(\ref{eq:sigma}), which gives the three-dimensional profile of the slip-face of the migrating dune as a function of time, can be now used in order to compute the sand flux along the dune axis. Avalanches occur along the slip-face in the direction that makes the smallest angle with the vertical direction, i.e. with the downward vector $-\hat{z}$. The plane of the slip-face is defined by this vector and the normal vector, $\hat{n} \equiv {\vec{N}}/||{\vec{N}}||$, where,
\begin{equation}
\begin{split}
\vec{N} & = \frac{\partial\sigma}{\partial x}\times \frac{\partial\sigma}{\partial y} = \\ & (1,0,-\tan(\theta_{c}))\times (0,1,D\omega\sin{(\omega y)}\tan{(\theta_{c})}) = \\
& (\tan(\theta_{c}),-D\omega\sin{(\omega y)}\tan{(\theta_{c})},1),
\end{split}
\end{equation}
and 
\begin{equation}
D=\frac{c\epsilon}{\Lambda\sqrt{({A_{0}})}}\exp({\Lambda t}).
\label{eq:D}
\end{equation} 
In this manner, the vector normal to the slip-face's surface is given by the equation,
\begin{equation}
\hat{n}=\frac{\overrightarrow{N}}{||\overrightarrow{N}||}=\frac{1}{C}(\sin(\theta_{c}),-D\omega\sin{(\omega y)}\sin{(\theta_{c})},\cos(\theta_{c})),
\end{equation}
where 
\begin{equation}
C=\sqrt{1+D^{2}\omega^{2}\sin^{2}{(\omega y)}\sin^{2}{(\theta_{c})}}.
\label{eq:C}
\end{equation}
So the direction of the flux is given by the normalized vector $\hat{f} = \overrightarrow{F}/||{\overrightarrow{F}}||$, where $\overrightarrow{F}$ is determined by a linear combination of $\hat{n}$ and $-\hat{z}$,
\begin{equation}
\overrightarrow{F}=\gamma\hat{n}-\beta\hat{z}. 
\label{eq:F}
\end{equation}
Since $\overrightarrow{F}$ is parallel to the surface of the slip-face, we conclude that $\overrightarrow{F} \cdot \hat{n}=0$. Combining this equation with Eq.~(\ref{eq:F}), we obtain a relation between $\gamma$ and $\beta$,
\begin{equation}
\gamma-\beta\hat{z} \cdot \hat{n}=0 \Longrightarrow \gamma=\beta\hat{z} \cdot \hat{n}=\frac{\beta\cos(\theta_{c})}{C}.
\end{equation}
In this manner, Eq.~(\ref{eq:F}) becomes,
\begin{equation}
\begin{split}
\overrightarrow{F} & =\frac{\beta}{C^{2}}(\sin(\theta_{c})\cos(\theta_{c}), \\ & -D\omega\sin{(\omega y)}\sin{(\theta_{c})}\cos(\theta_{c}),\cos^{2}(\theta_{c})-C^{2}),
\end{split}
\end{equation}
from which we obtain the normalized vector field of the flux of avalanches (see Fig.~\ref{fig:hygor}b).
\begin{equation}
\begin{split}
\hat{f}& =\frac{\overrightarrow{F}}{||\overrightarrow{F}||}=\frac{1}{C\sqrt{{C^{2}-\cos^{2}({\theta_{c}})}}}(\cos(\theta_{c})\sin(\theta_{c}), \\ & -D\omega\sin{(\omega y)}\cos{(\theta_{c})}\sin(\theta_{c}),\cos^{2}(\theta_{c})-C^{2}).
\end{split}
\label{eq:f}
\end{equation}

{\em{Condition for avalanches}} --- Avalanches will occur in the direction of $\hat{f}$ (Eq.~(\ref{eq:f})) wherever the angle $\theta$ defined by $\hat{f}$ and the $xy$-plane exceeds the angle of repose, $\theta_{c}=34^{o}$. The angle $\alpha$ between $\hat{f}$ and the vertical is given by the equation,
\begin{equation}
\begin{split}
\cos(\alpha) & = \overrightarrow{f} \cdot (-\hat{z}) \\
& = -\frac{\cos^{2}(\theta_{c})-C^{2}}{C\sqrt{{C^{2}-\cos^{2}({\theta_{c}})}}} = \frac{\sqrt{{C^{2}-\cos^{2}({\theta_{c}})}}}{C}. 
\end{split}
\end{equation}
In fact, the critical angle is measured between the plane of the slip-face and the horizontal plane, i.e. $\theta=\frac{\pi}{2}-\alpha$, such that,
\begin{equation}
\sin(\theta)= \frac{\sqrt{{C^{2}-\cos^{2}({\theta_{c}})}}}{C}
\end{equation}
Considering $\theta=\theta_{c}+\eta$, where $\eta$ is of second order in $\epsilon$, we can then use the following approximation: 
\begin{equation}
\sin(\theta)=\sin(\theta_{c})\cos(\eta)+\sin(\eta)\cos(\theta_{c}) \approx\eta\cos(\theta_{c}).
\label{eq:eta}
\end{equation}
{\em{Calculation of the sand flux}} --- Next we apply the continuity equation to calculate the transport of sand through avalanches. The mass $M(y,t)$ of an infinitesimally small slice of the dune width $dy$ and area $A(y,t)$ evolves in time as sand is transported along the axis of the transverse dune, i.e. in the direction $\hat{y}$, which is perpendicular to the surface area of any longitudinal slice.  Let us call $\overrightarrow{u}$ the velocity vector of the sand flow along the slip-face. From mass conservation, we obtain,
\begin{equation}
\begin{split}
M(y,t+dt) & = M(y,t)-\{\rho\overrightarrow{u}(y+\frac{dy}{2},t) \cdot \hat{y}A(y,t)dt \\ 
& +\rho\overrightarrow{u}(y-\frac{dy}{2},t) \cdot (-\hat{y})A(y,t)dt\}.
\end{split}
\end{equation}
Since $M(y,t)\approx \rho A(y,t)dy$, we can write,
\begin{equation}
\begin{split}
\rho\frac{\partial A(y,t)}{\partial t}dy &= -\{\rho(\overrightarrow{u}(y,t)+\frac{\partial\overrightarrow{u}(y,t)}{\partial y}\frac{dy}{2}).\hat{y}A(y,t) \\
& -\rho(\overrightarrow{u}(y,t) -\frac{\partial\overrightarrow{u}(y,t)}{\partial y}\frac{dy}{2}) \cdot \hat{y}A(y,t)\}.
\end{split}
\end{equation}
Thus,
\begin{equation}
\rho\frac{\partial A(y,t)}{\partial t}=-\rho\frac{\partial\overrightarrow{u}(y,t)}{\partial y} \cdot \hat{y}A(y,t),
\end{equation}
which leads to the following equation,
\begin{equation}
\frac{\partial A(y,t)}{\partial t}=-A(y,t)\frac{\partial u_{y}(y,t)}{\partial y},
\label{eq:continuity}
\end{equation}
where $u_y$ is the $y$-component of the velocity $\overrightarrow{u}$ defining the sand flux. The sand flux is proportional to the difference between the slope in the $y$-direction with the critical angle, whereas the direction is given by the vector $\hat{f}$. By defining the velocity vector of the sand flux as $\overrightarrow{u}=B\eta\hat{f}$, where $B$ is the characteristic velocity of the avalanches along the slip-face, and using Eq.(15), we obtain,
\begin{equation}
u_{y}(y,t)=B\eta\hat{f} \cdot \hat{y}=B\eta\left(\frac{-D\omega\sin{(\omega y)}\cos{(\theta_{c})}\sin(\theta_{c})}{C\sqrt{{C^{2}-\cos^{2}({\theta_{c}})}}}\right).
\end{equation}
From Eq.~(\ref{eq:eta}),
\begin{equation}
\begin{split}
u_{y}(y,t) & = \frac{B}{\cos(\theta_{c})}\frac{\sqrt{{C^{2}-\cos^{2}({\theta_{c}})}}}{C} \\ & \cdot \left(\frac{-D\omega\sin{(\omega y)}\cos{(\theta_{c})}\sin(\theta_{c})}{C\sqrt{{C^{2}-\cos^{2}({\theta_{c}})}}}\right) \\
& = -B\frac{D\omega\sin{(\omega y)}\sin(\theta_{c})}{C^{2}},
\end{split}
\end{equation}
using the values of $C$ and $D$ given by Eqs.~(\ref{eq:C}) and ({\ref{eq:D}}), respectively, and noting that $C \approx \sqrt{(1+O(\epsilon^{2}))}$, results in,
\begin{equation}
u_{y}(y,t)=-B\frac{c\epsilon\exp({\Lambda t})\omega\sin{(\omega y)}\sin(\theta_{c})}{\Lambda\sqrt{{A_{0}}}}.
\end{equation}
In this manner,
\begin{equation}
\frac{\partial u_{y}(y,t)}{\partial y}=-B\frac{c\epsilon\exp({\Lambda t})\omega^{2}\cos{(\omega y)}\sin(\theta_{c})}{\Lambda\sqrt{{A_{0}}}}. 
\label{eq:du_dy}
\end{equation}
Combining Eqs.~(\ref{eq:Area_pert}), (\ref{eq:continuity}) and (\ref{eq:du_dy}), leads to,
\begin{equation}
\begin{split}
A_{0}\Lambda\epsilon\exp({\Lambda t})\cos{(\omega y)} & = A_{0}(1+\epsilon\exp({\Lambda t})\cos{(\omega y)})B \\ & \cdot \frac{c\epsilon\exp({\Lambda t})\omega^{2}\cos{(\omega y)}\sin(\theta_{c})}{\Lambda\sqrt{{A_{0}}}}.
\end{split}
\end{equation}
Neglecting the terms of second order in $\epsilon$, we obtain,
\begin{equation}
A_{0}\Lambda\epsilon\exp({\Lambda t})\cos{(\omega y)}=A_{0}B\frac{c\epsilon\exp({\Lambda t})\omega^{2}\cos{(\omega y)}\sin(\theta_{c})}{\Lambda\sqrt{{A_{0}}}},
\end{equation}
and the growth rate of the perturbation can be finally written as,
\begin{equation}
\Lambda=\pm\omega\sqrt{\frac{cB\sin(\theta_{c})}{\sqrt{A_{0}}}}.
\label{eq:Lambda}
\end{equation}
Since the positive root in Eq.~(\ref{eq:Lambda}) dominates, the perturbation should always grow in time. Therefore, the linear
stability analysis corroborates the results from previous numerical simulations \citep{Parteli_et_al_2011} indicating that transverse dunes are unstable with respect to any perturbations along their axis. \\

\section{Concluding remarks}

In conclusion, we have shown, by means of a linear stability analysis of the transverse dune, that any perturbation along the axis of a transverse dune amplifies due to a combined effect of the following two main factors: the lateral transport of sand through avalanches along the dune's slip-face, and the scaling of dune migration velocity with the inverse of the dune height. It is important to emphasize that the growth rate estimated in Eq. (\ref{eq:Lambda}) refers only to the initial stage of the dune evolution, where the perturbation can be considered small. As the perturbation increases, nonlinear effects become important. Numerical simulations showed that, in this regime, the growth rate of the perturbations scales with the inverse of dune's turnover time, which is the time the dune needs to migrate a distance equivalent to its width \citep{Parteli_et_al_2011}. It would be interesting to conduct a theoretical study of stability where nonlinear effects are considered in order to compare the results with previously reported numerical simulations \citep{Parteli_et_al_2011}.

The results of the present paper apply to a transverse dune which is moving isolated on the bedrock in an area where the wind is strictly unidirectional. Variations in wind direction may enhance the transverse instability and accelerate the decaying process of a transverse dune into barchans. As shown previously, the surface of a large barchan may develop small superimposed structures --- the so-called sand-wave-instabilities \citep{Elbelrhiti_et_al_2005} --- when subjected to a secondary wind direction that makes a small angle with the primary wind (the one which forms the dune). The secondary bedforms developing on the dune's surface, which in fact are small dunes with wavelength of the order of the minimal dune size, migrate along the dune's crest thereafter leaving through the limbs. In this manner, large amounts of sand can be carried away from large desert dunes due to the occurence of secondary winds obliquely to the direction of motion. The role of wind trend variations for growth rate of the transverse instabilities is an open question that remains to be investigated in the future. Furthermore, spatial variations in the influx along the direction perpendicular to the wind, e.g. due to  the presence of dunes upwind, should also contribute to break the dune, and must be thus considered in a more detailed study of the transverse instability.

Our calculations elucidate the observation that, when the wind blows nearly from the same direction, the dominant dune shape moving on the bedrock is the barchan, and not the transverse dune \citep{Reffet_et_al_2010,Parteli_et_al_2011}. In fact, the present linear stability analysis shows that real transverse dunes moving under unidirectional winds are always unstable, independently of the amount of sand on the ground. Our calculations do not distinguish between high and low sand availability, i.e., the physical mechanisms driving the transverse instability exist also when transverse dunes evolve in areas of high sand availability, as for example on a deep sand bed. Indeed, real transverse dunes are never straight, but always display a characteristic sinusoidal shape, which is a fingerprint of the intrinsic instability reported here. However, a transverse dune can only fully break apart into a chain of barchans if it is moving on bedrock in an area with low sand supply \citep{Parteli_et_al_2011}. Furthermore, since the linear stability analysis cannot go beyond small deviations from the perfect translational invariance, it can not discriminate between a barchanoidal chain, a decomposition into individual barchans, or a wavy shaped transverse dune with eventually propagating waves. The study of the long-term evolution of the transverse instability when the dune is on a sand bed, or when the ground is wet or vegetated, might be tackled in the future with large scale numerical simulations \citep{Parteli_et_al_2011}. 

\section*{Acknowledgments}
We thank Heitor Credidio, Leandro Jader and Thomas Pähtz for discussions, and Sylvain Courrech du Pont for the images of his experiment. We also thank the Brazilian agencies CNPq, CAPES, FUNCAP, FINEP, and the CNPq/FUNCAP Pronex grant, as well as the INST-SC and the ETH Grant ETH-10 09-2 for financial support.

\bibliographystyle{model1-num-names}

\begin{thebibliography}{00}

\bibitem[Bagnold(1941)]{Bagnold_1941} Bagnold, R.A., 1941. Physics of blown sand and desert dunes. Springer, London, 289 pp.

\bibitem[Cutts and Smith(1973)]{Cutts_and_Smith_1973} Cutts, J.A., Smith, R.S.U., 1973. Eolian Deposits and Dunes on Mars. Journal of Geophysical Research 78, 4139-4154.

\bibitem[McCauley(1973)]{McCauley_1973} McCauley, J.F., 1973. Mariner 9 Evidence for Wind Erosion in the Equatorial and Mid-Latitude Regions of Mars. Journal of Geophysical Rsearch 78, 4123-4137.

\bibitem[Greeley et al.(1992)]{Greeley_et_al_1992} Greeley, R., Arvidson, R.E., Elachi, C., Geringer, M.A., Plaut, J.J., Stephen Saunders, R., Schubert, G., Stofan, E.R., Thouvenot, E.J.P., Wall, S.D., Weitz, C.M., 1992. Aeolian features on Venus: Preliminary Magellan results. Journal of Geophysical Research 97, 13319-13345.

\bibitem[Lorenz et al.(2006)]{Lorenz_et_al_2006} Lorenz, R.D., Wall, S., Radebaugh, J., Boubin, G., Reffet, E., Janssen, M., Stofan, E., Lopes, R., Kirk, R., Elachi, C., Lunine, J., Mitchell, K., Paganelli, F., Soderblom, L., Wood, C., Wye, L., Zebker, H., Anderson, Y., Ostro, S., Allison, M., Boehmer, R., Callahan, P., Encrenaz, P., Ori1, G.G., Francescetti, G., Gim, Y., Hamilton, G., Hensley, S., Johnson, W., Kelleher, K., Muhleman, D., Picardi, G., Posa, F., Roth, L., Seu, R., Shaffer, S., Stiles, B., Vetrella, S., Flamini, E., West, R., 2006. The Sand Seas of Titan: Cassini RADAR Observations of Longitudinal Dunes. Science 312, 724-727.

\bibitem[Fourri\`ere et al.(2010)]{Fourriere_et_al_2010} Fourri\`ere, A., Claudin, P., Andreotti, B., 2010. Bedforms in a turbulent stream: formation of ripples by primary linear instability and of dunes by nonlinear pattern coarsening. J. Fluid Mech. 649, 287-328.

\bibitem[Bourke et al.(2010)]{Bourke_et_al_2010} Bourke, M.C, Lancaster, N., Fenton, L.K., Parteli, E.J.R., Zimbelman, J.R., Radebaugh, J., 2010. Extraterrestrial dunes: An introduction to the special issue on planetary dune systems. Geomorphology 121, 1-14.

\bibitem[Pye and Tsoar(1991)]{Pye_and_Tsoar_1991} Pye, K., Tsoar, H., 1991. Aeolian sand and sand dunes. Springer, London., 289 pp.

\bibitem[Wiggs(2001)]{Wiggs_2001} Wiggs, G.F.S., 2001. Desert dune processes and dynamics. Progress in Physical Geography 25, 53-79.

\bibitem[Hersen et al.(2002)]{Hersen_et_al_2002} Hersen, P., Douady, S., Andreotti, B., 2002. Relevant Length Scale of Barchan Dunes. Physical Review Letters 89, 264301.

\bibitem[Elbelrhiti et al.(2005)]{Elbelrhiti_et_al_2005} Elbelrhiti, H., Claudin, P., Andreotti, B., 2005. Field evidence for surface-wave-induced instability of sand dunes. Nature, 437, 720-723.

\bibitem[Livingstone et al.(2007)]{Livingstone_et_al_2007} Livingstone, I., Wiggs, G.F.S., Weaver, C.M., 2007. Geomorphology of desert sand dunes: A review of recent progress. Earth-Science Reviews 80, 239-257.

\bibitem[Andreotti et al.(2009)]{Andreotti_et_al_2009} Andreotti, B., Claudin, P., Fourri\`ere, A., Ould-Kadddour, F., Murray, B., 2009. Giant aeolian dune size determined by the average depth of the atmospheric boundary layer. Nature, 457, 1120-1123.

\bibitem[Andreotti et al.(2010)]{Andreotti_et_al_2010} Andreotti, B., Claudin, P., Pouliquen, O., 2010. Measurements of the aeolian sand transport saturation length. Geomorphology 123, 343-348.

\bibitem[Reffet et al.(2010)]{Reffet_et_al_2010} Reffet, E., Courrech du Pont, S., Hersen, P., Douady, S., 2010. Formation and stability of transverse and longitudinal sand dunes. Geology 38, 491-494.

\bibitem[Werner(1995)]{Werner_1995} Werner, B.T., 1995. Eolian dunes: Computer simulation and attractor interpretation. Geology, 23, 1107-1110.

\bibitem[Nishimori et al.(1998)]{Nishimori_et_al_1998} Nishimori, H., Yamasaki, M., Andersen, K.H., 1998. A simple model for the various pattern dynamics of dunes. Int. J. Mod. Phys. B 12, 257-272.

\bibitem[Sauermann et al.(2001)]{Sauermann_et_al_2001} Sauermann, G., Kroy, K., Herrmann, H.J., 2001. A continuum saltation model for sand dunes. Physical Review E 64, 31305.

\bibitem[Kroy et al.(2002)]{Kroy_et_al_2002} Kroy, K., Sauermann, G., Herrmann, H.J., 2002. Minimal model for aeolian sand dunes. Physical Review E 66, 031302.

\bibitem[Herrmann et al.(2008)]{Herrmann_et_al_2008} Herrmann, H.J., Dur\'an, O., Parteli, E.J.R., Schatz, V., 2008. Vegetation and induration as sand dunes stabilizers. Journal of Coastal Research 24, 1357-1368.

\bibitem[Narteau et al.(2009)]{Narteau_et_al_2009} Narteau, C., Zhang, D., Rozier, O., Claudin, P., 2009. Setting the length and time scales of a cellular automaton dune model from the analysis of superimposed bed forms. J. Geophys. Res.114, F03006.

\bibitem[Parteli et al.(2009)]{Parteli_et_al_2009} Parteli, E.J.R., Dur\'an, P., Tsoar, H., Schw\"ammle, V., Herrmann, H.J., 2009. Dune formation under bimodal winds. Proc. Nat. Acad. Sci. 106, 22085-22089.

\bibitem[Andreotti(2004)]{Andreotti_2004} Andreotti, B., 2004. A two-species model of aeolian sand transport. Journal of Fluid Mechanics 510, 47-70.

\bibitem[Almeida et al.(2006)]{Almeida_et_al_2006} Almeida, M.P., Andrade Jr., J.S., Herrmann, H.J., 2006. Aeolian transport layer. Physical Review Letters 96, 018001.

\bibitem[Almeida et al.(2008)]{Almeida_et_al_2008} Almeida, M.P., Parteli, E.J.R., Andrade Jr., J.S., Herrmann, H.J., 2008. Giant saltation on Mars. Proc. Natl. Acad. Sci. 105, 6222-6226.

\bibitem[Kok and Renno(2009)]{Kok_and_Renno_2009} Kok, J.F., Renno, N.O., 2009. A comprehensive numerical model of steady-state saltation (COMSALT), Journal of Geophysical Research 114, D17204.

\bibitem[Wasson and Hyde(1983)]{Wasson_and_Hyde_1983} Wasson, R.J., Hyde, R., 1983. Factors determining desert dune type. Nature 304, 337-339.

\bibitem[Finkel(1959)]{Finkel_1959} Finkel, H.J., 1959. The barchans of southern Peru. Journal of Geology 67, 614-647.

\bibitem[Long and Sharp(1964)]{Long_and_Sharp_1964} Long, J.T., Sharp, R.P., 1964. Barchan-dune movement in Imperial Valley, California. Geol. Soc. Am. Bull. 75, 149-156.

\bibitem[Embabi and Ashour(1993)]{Embabi_and_Ashour_1993} Embabi, N.S., Ashour, M.M., 1993. Barchan dunes in Qatar. Journal of Arid Environments 25, 49-69.

\bibitem[Hesp and Hastings(1998)]{Hesp_and_Hastings_1998} Hesp, P.A., Hastings, K., 1998. Width, height and slope relationships and aerodynamic maintenance of barchans. Geomorphology 22, 193-204.

\bibitem[Sauermann et al.(2000)]{Sauermann_et_al_2000} Sauermann, G., Rognon, P., Poliakov, A., Herrmann, H.J., 2000. The shape of the barchan dunes of southern Morocco. Geomorphology 36, 47-62.

\bibitem[Sauermann et al.(2003)]{Sauermann_et_al_2003} Sauermann, G., Andrade Jr., J.S., Maia, L.P., Costa, U.M.S., Ara\'ujo, A.D., Herrmann, H.J., 2003. Wind velocity and sand transport on a barchan dune. Geomorphology 54, 245-255.

\bibitem[Bourke and Goudie(2009)]{Bourke_and_Goudie_2009} Bourke, M.C, Goudie, A.S., 2009. Varieties of barchan dunes in the Namib Desert and on Mars. Aeolian Research 1, 45-54. 

\bibitem[Mulligan(1988)]{Mulligan_1988} Mulligan, K.R., 1988. Velocity profiles measured on the windward slope of a transverse dune. Earth Surface Processes and Landforms 13, 573-582.

\bibitem[McKenna Neuman et al.(2000)]{McKenna_Neuman_et_al_2000} McKenna Neuman, C., Lancaster, N., Nickling, W.G., 2000. The effect of unsteady winds on sediment transport on the stoss slope of a transverse dune, Silver Peak, NV, USA. Sedimentology 47, 211-226.

\bibitem[Walker and Nickling(2002)]{Walker_and_Nickling_2002} Walker, I.J., Nickling, W.G., 2002. Dynamics of secondary airflow and sediment transport over and in the lee of transverse dunes. Progress in Physical Geography 26, 47-75.

\bibitem[Parteli et al.(2006)]{Parteli_et_al_2006} Parteli, E.J.R., Schw\"ammle, V., Herrmann, H.J., Monteiro, L.H.U., Maia, L.P., 2006. Profile measurement and simulation of a transverse dune field in the Lencois Maranhenses. Geomorphology 81, 29-42.

\bibitem[Lancaster(1995)]{Lancaster_1995} Lancaster, N., 1995. Geomorphology of desert dunes. Routledge, London, 290 pp.

\bibitem[Bourke et al.(2004)]{Bourke_et_al_2004} Bourke, M.C, Balme, M., Zimbelman, J.R., 2004. A comparative analysis of barchan dunes in the intra-crater dune fields and the North Polar Sand Sea, LPSC XXXV abst. 1453. 

\bibitem[Katsuki et al.(2005)]{Katsuki_et_al_2005} Katsuki, A., Kikuchi, M., Endo, N., 2005. Emergence of a Barchn Belt in a Unidirectional Flow: Experiment and Numerical Simulation. Journal of the Physical Society of Japan 74, 878-881.

\bibitem[Dur\'an et al.(2010)]{Duran_et_al_2010} Dur\'an, O., Parteli, E.J.R., Herrmann, H.J., 2010. A continuous model for sand dunes: Review, new developments and application to barchan dunes and barchan dune fields. Earth Surface Processes and Landforms 35, 1591-1600.

\bibitem[Luna et al.(2011)]{Luna_et_al_2011} Luna, M.C.M.M., Parteli, E.J.R., Dur\'an, O., Herrmann, H.J., 2011. Model for the genesis of coastal dune fields with vegetation. Geomorphology 129, 215-224.

\bibitem[Parteli et al.(2011)]{Parteli_et_al_2011} Parteli, E.J.R., Andrade Jr., J.S., Herrmann, H.J., 2011. Transverse instability of dunes. Physical Review Letters 107, 188001.

\bibitem[Kocurek et al.(1992)]{Kocurek_et_al_1992} Kocurek, G., Townsley, M., Yeh, E., Havholm, K., Sweet, M.L., 1992. Dune and dunefield development on Padre Island Texas, with implications for interdune deposition and water-table-controlled accumulation. J. Sediment. Petrol. 62, 622-635.

\bibitem[Kennedy(1969)]{Kennedy_1969} Kennedy, J.F., 1969. The formation of sediment ripples, dunes and antidunes. Annu. Rev. Fluid Mech. 1, 147-168.

\bibitem[Smith(1970)]{Smith_1970} Smith, J.D., 1970. Stability of a sand bed subjected to a shear flow at low Froude number. J. Geophys. Res. 75, 5928-5940.

\bibitem[Richards(1980)]{Richards_1980} Richards, K. J. 1980. The formation of ripples and dunes on an erodible bed. J. Fluid Mech. 99, 597-618.

\bibitem[McLean(1990)]{McLean_1990} McLean, S.R., 1990. The stability of ripples and dunes. Earth-Science Rev. 29, 131-144.

\bibitem[Stam(1997)]{Stam_1997} Stam, J.M.T., 1997. On the modelling of two-dimensional aeolian dunes. Sedimentology 44, 127-141.

\bibitem[van Dijk et al.(1999)]{van_Dijk_et_al_1999} van Dijk, P.M., Arens, S.M., van Boxel, J.H., 1999. Aeolian processes accross transverse dunes II: Modelling the sediment transport and profile development. Earth Surface Processes and Landforms 24, 319-333.

\bibitem[Andreotti et al.(2002)]{Andreotti_et_al_2002} Andreotti, B., Claudin, P., Douady, S., 2002. Selection of dune shapes and velocities. Part 2: A two-dimensional modelling. The European Physical Journal B 28, 341-352.

\bibitem[Schw\"ammle and Herrmann(2004)]{Schwaemmle_and_Herrmann_2004} Schw\"ammle, V., Herrmann, H.J., 2004. Modelling transverse dunes. Earth Surface Processes and Landforms 29, 769-784.

\bibitem[Anderson(1987)]{Anderson_1987} Anderson, R. 1987. A theoretical model for aeolian impact ripples. Sedimentology 34, 943-956.

\bibitem[Fowler(2001)]{Fowler_2001} Fowler, A.C., 2001. Dunes and drumlins. In: Geomorphological fluid mechanics, eds. A. Provenzale and N. Balmforth, pp. 430-454, Springer-Verlag, Berlin. 

\bibitem[Fowler(2002)]{Fowler_2002} Fowler, A.C., 2002. Evolution equations for dunes and drumlins. Revista de la Real Academia de Ciencias Exactas, Físicas y Naturales, Serie A. Mat. 96, 377-387. 

\bibitem[Herrmann et al.(2005)]{Herrmann_et_al_2005} Herrmann, H.J., Andrade Jr., J.S., Schatz, V., Sauermann, G., Parteli, E.J.R., 2005. Calculation of the separation streamlines of barchans and transverse dunes. Physica A 357, 44-49.

\bibitem[Allen(1974)]{Allen_1974} Allen, J.R.L., 1974. Reaction, relaxation and lag in natural sedimentary systems: General principles, examples and lessons. Earth Sci. Rev. 10, 263-342.

\bibitem[Hersen et al.(2004)]{Hersen_et_al_2004} Hersen, P., Andersen, H., Elbelrhiti, H., Andreotti, B., Claudin, P., Douady, S., 2004. Corridors of barchan dunes: Stability and size selection. Physical Review E 69, 011304.




\end{thebibliography}

\end{document}